\documentclass[aip,rsi,amsmath,amssymb,reprint,]{revtex4-1}

\usepackage{graphicx}
\usepackage{dcolumn}
\usepackage{bm}

\begin{document}

\title{InGaAs/InP single-photon detectors with 60\% detection efficiency at 1550 nm}

\author{Yu-Qiang Fang}
 \affiliation{Hefei National Laboratory for Physical Sciences at the Microscale and Department of Modern Physics, University of Science and Technology of China, Hefei 230026, China}
 \affiliation{CAS Center for Excellence in Quantum Information and Quantum Physics, University of Science and Technology of China, Hefei 230026, China}
\author{Wei Chen}
 \affiliation{China Electronics Technology Group Corporation No. 44 Research Institute, Chongqing 400060, China}
 \affiliation{Chongqing Key Laboratory of Core Optoelectronic Devices for Quantum Communication, Chongqing 400060, China}
\author{Tian-Hong Ao}
 \affiliation{China Electronics Technology Group Corporation No. 44 Research Institute, Chongqing 400060, China}
 \affiliation{Chongqing Key Laboratory of Core Optoelectronic Devices for Quantum Communication, Chongqing 400060, China}
\author{Cong Liu}
 \affiliation{China Electronics Technology Group Corporation No. 44 Research Institute, Chongqing 400060, China}
\author{Li Wang}
 \affiliation{China Electronics Technology Group Corporation No. 44 Research Institute, Chongqing 400060, China}
\author{Xin-Jiang Gao}
 \email{gaoxj@cetccq.com.cn}
 \affiliation{China Electronics Technology Group Corporation No. 44 Research Institute, Chongqing 400060, China}
 \affiliation{Chongqing Key Laboratory of Core Optoelectronic Devices for Quantum Communication, Chongqing 400060, China}
\author{Jun Zhang}
 \email{zhangjun@ustc.edu.cn}
 \affiliation{Hefei National Laboratory for Physical Sciences at the Microscale and Department of Modern Physics, University of Science and Technology of China, Hefei 230026, China}
 \affiliation{CAS Center for Excellence in Quantum Information and Quantum Physics, University of Science and Technology of China, Hefei 230026, China}
\author{Jian-Wei Pan}
 \affiliation{Hefei National Laboratory for Physical Sciences at the Microscale and Department of Modern Physics, University of Science and Technology of China, Hefei 230026, China}
 \affiliation{CAS Center for Excellence in Quantum Information and Quantum Physics, University of Science and Technology of China, Hefei 230026, China}

\date{\today}

\begin{abstract}
InGaAs/InP single-photon detectors (SPDs) are widely used for near-infrared photon counting in practical applications.
Photon detection efficiency (PDE) is one of the most important parameters for SPD characterization, and therefore increasing PDE consistently plays a central role in both industrial development and academic research.
Here we present the implementation of high-frequency gating InGaAs/InP SPD with a PDE as high as 60\% at 1550 nm.
On one hand, we optimize the structure design and device fabrication of InGaAs/InP single-photon avalanche diode with an additional dielectric-metal reflection layer to relatively increase the absorption efficiency of incident photons by $\sim$ 20\%.
On the other hand, we develop a monolithic readout circuit of weak avalanche extraction to minimize the parasitic capacitance for the suppression of the afterpulsing effect.
With 1.25 GHz sine wave gating and optimized gate amplitude and operation temperature, the SPD is characterized to reach a PDE of 60\% with a dark count rate (DCR) of 340 kcps.
For practical use, given 3 kcps DCR as a reference the PDE reaches $\sim$ 40\% PDE with an afterpulse probability of 5.5\%, which can significantly improve the performance for the near-infrared SPD based applications.
\end{abstract}

\maketitle

\section{Introduction}

Single-photon detectors (SPDs) are the most sensitive tools for weak light detection~\cite{SPD09,SPD11}.
In the near-infrared range, SPDs are extensively required in numerous applications such as quantum key distribution~\cite{QKD02}, lidar~\cite{Yu17}, and optical time-domain reflectometer~\cite{OTDR10}.
Currently, the practical technologies for infrared single-photon detection include superconducting nanowire single-photon detector (SNSPD)~\cite{SNSPD13,SNSPD17}, up-conversion single-photon detector~\cite{UCSPD05,Shentu13}, and InGaAs/InP single-photon detector~\cite{TMZ09,IJE11,Zhang15}.
Each SPD technology has its advantages and disadvantages.
For instance, InGaAs/InP SPDs are widely used in practical applications due to the features of small size, low cost, and easy operation, with a cost of relatively poor performance~\cite{Zhang15}.
The typical parameters to characterize InGaAs/InP SPDs include photon detection efficiency (PDE), dark count rate (DCR), afterpulse probability ($P_{ap}$), maximum count rate, and timing jitter, among which PDE is one of the most important figure of merits.
Commercially available InGaAs/InP SPDs exhibit a typical PDE of $\sim$ 20\% with a DCR of $\sim$ 500 cps at 1550 nm~\cite{IDQ,CTek}.
Recently, a miniaturized InGaAs/InP SPD with a PDE of 30\% and a DCR of 2 kcps has been reported~\cite{Jiang18}.
As a comparison, so far record values of PDE have been achieved as high as 93\%~\cite{SNSPD13} and 46\%~\cite{UCSPD05} for SNSPD and up-conversion SPD, respectively.

For an InGaAs/InP SPD consisting of a single-photon avalanche diode (SPAD) and a readout circuit, PDE is defined as the probability that the SPD system produces a desired output signal in response to an incident photon, which is determined by the coupling efficiency, the absorption efficiency, and the avalanche efficiency of InGaAs/InP SPAD~\cite{Zhang15}.
Given a pigtailed SPAD device, the coupling efficiency and the absorption efficiency are fixed, while the avalanche efficiency highly relies on the excess bias.
However, one cannot simply increase the excess bias to obtain high PDE since apart from PDE other parameters, particularly DCR and $P_{ap}$, are also related with the excess bias.
In practice, InGaAs/InP SPADs are operated either in free-running mode or gating mode~\cite{Zhang15}.
The free-running mode is suited for asynchronous single-photon detection, and so far many approaches for free-running operation have been reported~\cite{Rarity00,TSG07,Liu08,Zhang09,WIB09,Itzler09,Yan12,KWL14,Yu18}.
Nevertheless, the long hold-off time has to be applied to suppress the afterpulsing effect in such mode, resulting in a limited count rate for applications.
High-frequency gating, including self-differencing~\cite{SD07,SD09,SD10,SD15} and sine wave gating (SWG)~\cite{SWG06,Zhang09-2,Liang12,SWG12,SWG13,Jiang17,Jiang18}, is an effective technique to suppress the afterpulsing effect so that the count rate of SPD can be drastically increased.
In such a scheme, the gating time is limited to a few hundreds picoseconds.
As a result, the charge carrier quantity and thus the afterpulsing effect can be considerably suppressed, and meanwhile, the avalanche signals become very weak that may bring technology challenge to design readout circuit~\cite{Zhang15}.

In this paper, we present the implementation of 1.25 GHz sine wave gating InGaAs/InP SPDs with a PDE up to 60\% at 1550 nm.
We perform the structure design optimization of InGaAs/InP SPADs by adding a dielectric-metal reflection layer, and calibrated results on the fabricated sample SPAD devices show that the reflectivity of dielectric-metal reflection layer to the incident photons at 1550 nm reaches $\sim$ 80\%.
Further, we design and fabricate a monolithically integrated readout circuit (MIRC) dedicated to extracting weak avalanches for the scheme of 1.25 GHz sine wave gating.
By optimizing the gate amplitude and operation temperature, the InGaAs/InP SPDs are characterized to achieve PDE of 60.1\% at 300 K with a DCR of 340 kcps and a $P_{ap}$ of 14.8\% . For practical use, the SPD exhibits significantly better performance than commercial products, e.g., $\sim$ 40\% PDE with 3 kcps DCR and 5.5\% $P_{ap}$ at 253 K.

\section{SPAD design and fabrication}

\begin{figure}[htbp]
\centering
\includegraphics[width=\linewidth]{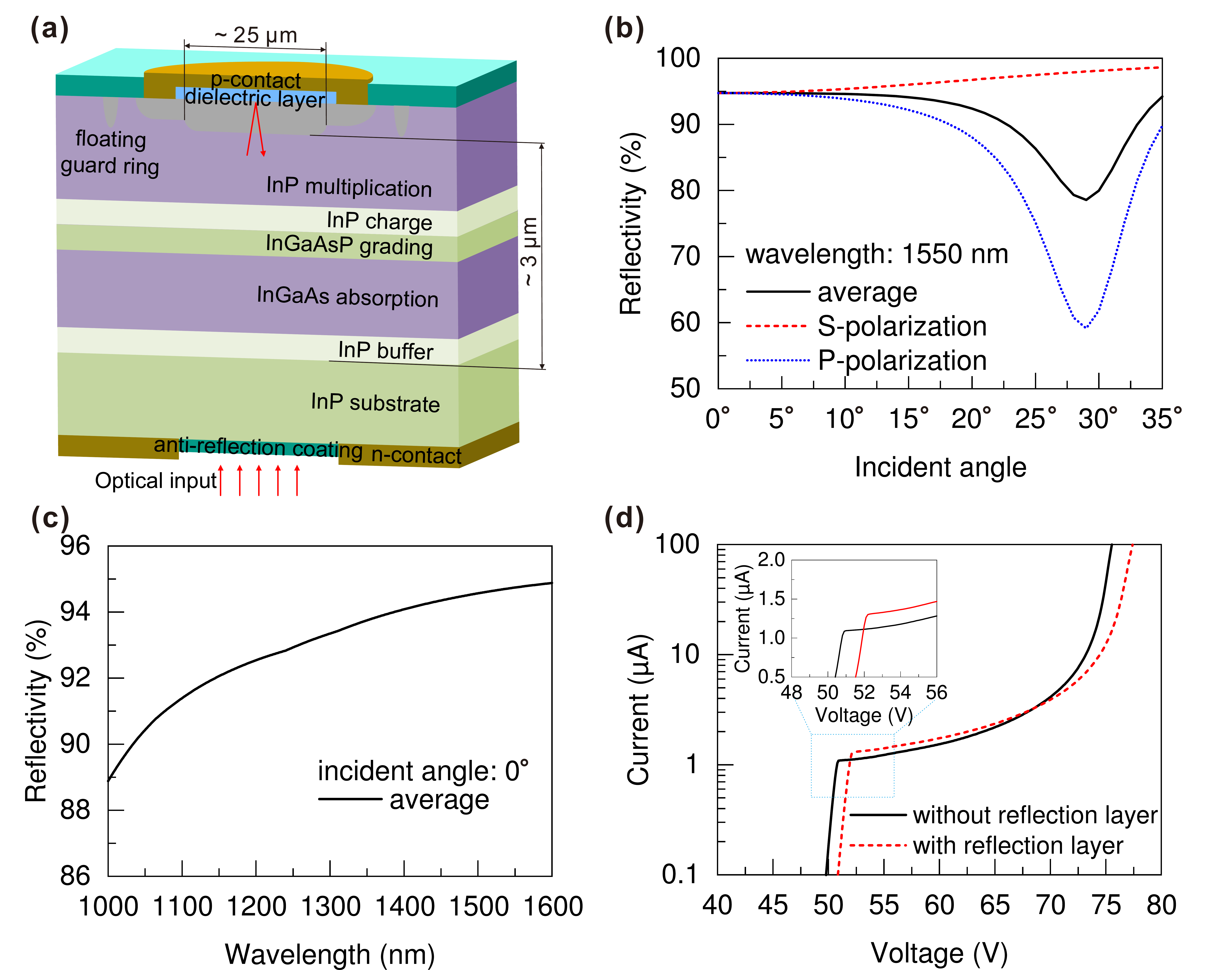}
\caption{InGaAs/InP SPAD design. (a) Schematic diagram of SPAD SAGCM structure with a dielectric-metal reflection layer on the top. (b) The reflectivity of S-polarization and P-polarization components of light as a function of the incident angle. (c) The average reflectivity as a function of the incident light wavelength at 0$^\circ$ incident angle. (d) Comparison of typical photocurrent-voltage (I-V) curves of SPAD chips with or without the reflection layer.}
\label{fig1}
\end{figure}

Following the approach~\cite{Ma16}, we perform semiconductor parameter design and optimization of InGaAs/InP SPAD based on a separate absorption, grading, charge, and multiplication (SAGCM) structure with a stepped PN junction~\cite{IBH07}, with an active area diameter of $\sim$ 25 $\mu$m, as illustrated in Fig.~\ref{fig1}a.
Specifically, compared with our previous design and fabrication~\cite{Jiang18}, a dielectric-metal reflection layer with high reflectivity~\cite{LBM09} is added to enhance the absorption efficiency of incident photons.
Once incident photons pass through the absorption layer from the substrate region, the unabsorbed photons are reflected by the reflection layer and reabsorbed during the absorption layer, which can effectively increase the overall absorption efficiency.

The dielectric-metal reflection layer, consisting of a dielectric layer and a metal layer, is located on the top of the photosensitive region.
The dielectric layer lies between the metal layer and the InP multiplication region.
The material of dielectric layer is SiO$_2$, and the thickness of the dielectric layer is 220 $\sim$ 230 nm.
The reflectivity is enhanced due to the interference of reflected light between the semiconductor-dielectric interface and the dielectric-metal interface.
We calculate the relationship between the reflectivity of the reflection layer and the angle of incident light at 1550 nm, and further optimize the thickness of the dielectric layer film.
Fig.~\ref{fig1}b plots the reflectivities of S-polarization and P-polarization components of the incident light as a function of incident angles, given that the angle perpendicular to the reflection layer is defined as 0$^\circ$.
At 0$^\circ$ incident angle, the reflectivities of S-polarization and P-polarization components are the same, and the average reflectivity reaches the largest value.
Further, we simulate the relationship between the average reflectivity and the light wavelength given 0$^\circ$ incident angle, as plotted in Fig.~\ref{fig1}c.
The simulation results show that the optimal reflectivity of the dielectric-metal reflection layer to the incident photon of 1550 nm can reach 95\%.

Apart from the reflection layer, the parameters of other layers are also optimized.
The electric field distributions in the absorption and multiplication layers are regulated by tuning the doping concentration in the grading and charge layers.
The electric field strength in the absorption layer is regulated in the range of 1 $\sim$ 1.4$\times 10^5$ V/cm to guarantee the saturation drift rate of the photogenerated carriers while avoiding avalanche breakdown.
The thickness of the multiplication layer is regulated in the range of 1 $\sim$ 1.5 $\mu$m to guarantee enough electric field strength for avalanche breakdown while reducing the carrier transit time and thus the timing jitter.
The total thickness of the depletion region is $\sim$ 3 $\mu$m.

The InGaAs/InP SPAD chips are fabricated via an epitaxial process of the metal-organic chemical vapor deposition (MOCVD).
During the fabrication, the background impurity concentration of epitaxial materials is reduced within the range of $2\times10^{14}$ $cm^{-3}$ and the surface charge density in the charge layer is controlled with an accuracy of 5\%.
The depth of PN junction is regulated with an accuracy of $\pm$50 nm via double-diffusion process technology using a Zn source~\cite{Jiang18}.
The p-contact region, located on the top surface of the stepped Zn-doped region, is coaxial with the photosensitive region, which can achieve high reflectivity due to the reflection layer while maintaining the contact resistance as small as possible.

In order to verify the effectiveness of the reflection layer, we design and fabricate two kinds of InGaAs/InP sample SPADs with and without the reflection layer, respectively. Apart from the reflection layer, all the other structure parameters in the two cases are exactly the same.
Fig.~\ref{fig1}d plots the measured photocurrent-voltage (I-V) curves of sample SPADs in two cases, in response to incident light at 1550 nm.
The photocurrents at the turning point of reach-through voltage in two cases are suited to calculate the reflectivity of the reflection layer.
The small offset of reach-through voltages in two cases is due to the parameter fluctuations during the fabrication process.
As shown in the inset of Fig.~\ref{fig1}d, under the same light illumination the photocurrents at the reach-through voltages are 1.30 $\mu$A and 1.09 $\mu$A with and without the reflection layer, respectively, which clearly indicates that the absorption efficiency and thus PDE can be relatively increased by $\sim$ 20\%
with the help of adding reflection layer. Further, according to the measured photocurrent, light intensity, and absorption layer thickness, the reflectivity of the reflection layer is calculated to be $\sim$ 80\%, which is less than the optimal simulation result of 95\%. Such difference is primarily contributed by two factors, i.e., the imperfect flatness of the dielectric layer during the fabrication process, and incident angle deviation from 0$^\circ$ induced by the coupling lens.

\section{SPD system and characterization}

\begin{figure}[htbp]
\centering
\includegraphics[width=\linewidth]{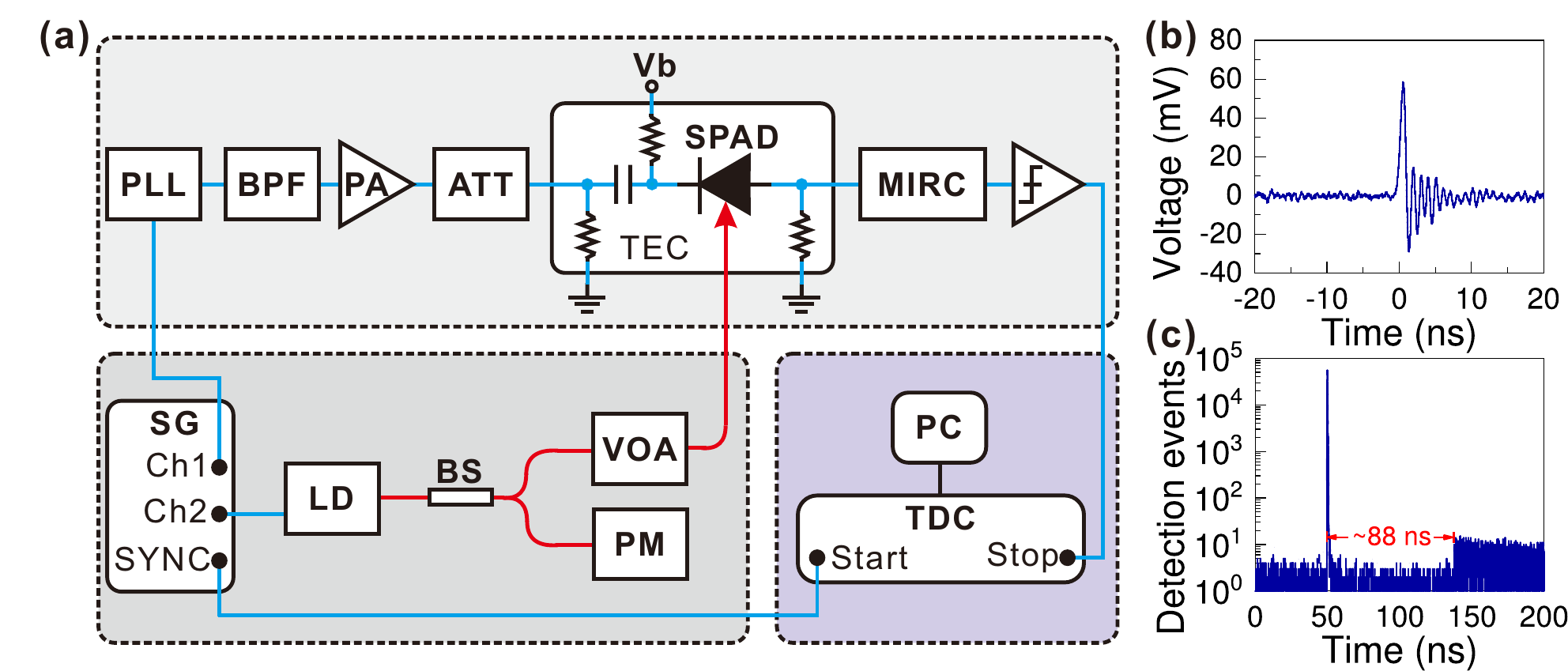}
\caption{(a) Experimental setup for SPD characterization. (b) A typical avalanche signal captured at the output of MIRC. (c) A typical detection event distribution measured by a TDC with an intrinsic deadtime of $\sim$ 88 ns.}
\label{fig2}
\end{figure}

The fabricated InGaAs/InP sample SPADs are operated in the high-frequency sine wave gating mode~\cite{Zhang15}.
The schematic diagram in Fig.~\ref{fig2}a illustrates the SPD module, calibration system, and measurement setup.
One channel (Ch1) of the signal generator (SG, Keysight 81150A) generates a 10 MHz clock, which is used as the reference signal of the phase-locked loop (PLL) to generate an initial 1.25 GHz square-wave signal.
After passing through a band-pass filter (BPF) with a center frequency of 1.2 GHz, the 1.25 GHz square-wave signal is transformed into a 1.25 GHz sine wave gate.
The sine wave gate is further regulated with a narrow-band radio frequency power amplifier (PA) and a variable attenuator (ATT), and then is alternating current (AC) coupled to the cathode of SPAD.
The amplitude of the sine wave gate can be amplified up to 30 V of $V_{pp}$ without distortion and be adjusted continuously.
In such SWG scheme, due to the pure frequency spectrum characteristic of sine waves, the capacitive response signals of SPAD are only composed of sine waves with the fundamental frequency and higher order harmonics, which can be easily eliminated by band-stop filters (BSFs)~\cite{SWG06,Zhang09-2,Liang12} or low-pass filters (LPFs)~\cite{SWG12,Jiang17,Jiang18}.

In the SPD system, we design a MIRC with a size of 15 mm $\times$ 15 mm to extract weak avalanche signals~\cite{Jiang17}.
MIRC integrates two LPFs and a two-stage radio frequency low-noise amplifier (LNA), in which each LPF has a cutoff frequency of $\sim$ 1 GHz and a rejection ratio of $\sim$ 60 dB at 1.25 GHz and the LNA has a gain of $\sim$ 40 dB below 1 GHz.
After testing with a network analyzer, MIRC exhibits radio frequency performance with a gain of $\sim$ 40 dB below 1 GHz and a rejection ratio of $\sim$ 80 dB at 1.25 GHz~\cite{Jiang17}.
MIRC is fabricated using the technology of low temperature co-fired ceramics, which is a standard printed circuit board like manufacturing process with multi-layer ceramic dielectric tapes and screen printing of conductors materials, e.g., silver, and is thus well suited for the integration of radio frequency components.
Using MIRC instead of conventional readout circuits composed of discrete electronic components can minimize the parasitic capacitance around the SPAD, which, therefore, substantially reduces the afterpusling effect.
A typical avalanche signal at the output of MIRC is captured by a high-speed oscilloscope (Keysight MSOS804A), as plotted in Fig.~\ref{fig2}b.
The rising edge of the avalanche signal is slowed down by the LPFs inside the MIRC, which contributes to the timing jitter of the SPD system.
The output signals of MIRC are further discriminated by a high-speed comparator.
The threshold of the comparator is chosen to be the value as small as $\sim$ 15 mV to ensure that no signals are discriminated given the bias below the breakdown voltage. The intrinsic deadtime of the SPD system is limited by the discrimination and monostable circuits, i.e., $\sim$ 3 ns.

We then follow the single-photon calibration scheme~\cite{Zhang15} to characterize the primary parameters of InGaAs/InP SPDs.
The second channel (Ch2) of SG generates a clock of 625 kHz to drive a picosecond pulsed laser diode (LD, QuantumCTek QCL-102).
The optical pulses emitted from the LD at 1550 nm with a full width at half maximum (FWHM) of 50 ps pass through a 99:1 beam splitter (BS), in which the 99\% port is monitored by a power meter (PM, EXFO IQS-1600) and the 1\% port is connected with a variable optical attenuator (VOA, EXFO IQS-3150).
The intensity of laser pulses is attenuated down to a level of mean photon number per pulse of 1.
The synchronized signals of Ch2 and the output signals from the discriminator are fed into a time-to-digital converter (TDC, PicoQuant picoharp 300) with an intrinsic deadtime of $\sim$ 88 ns as ``start'' and ``stop'', respectively.
TDC performs a timing tag for each detection event and transmits data to a personal computer (PC).
Fig.~\ref{fig2}c plots a typical histogram of detection event distribution.

With such measurement settings, the primary parameters of InGaAs/InP SPDs can be calculated.
Considering the Poisson distribution of laser source, PDE can be calculated as~\cite{Zhang15}
\begin{equation}\label{eq1}
PDE = \frac{1}{\mu}ln\frac{1-R_d/f_g}{1-R_{ph}/f_l},
\end{equation}
where $\mu$ is the mean photon number per laser pulse, $R_d$ is the measured count rates without laser illumination, $f_g$ is the gating frequency, $f_l$ is the laser repetition frequency, and $R_{ph}$ is the photon detection count rate with laser illumination, i.e., the coincidence rate between detections and laser pulses after subtracting the DCR contribution.
The normalized DCR, which is usually used for fair comparison in different conditions, can be calculated as $R_d/(f_gt_w)\times10^9$,
where $t_w$ is the effective gating width~\cite{Zhang15}. From the detection event distribution, $P_{ap}$ can be calculated as~\cite{Liang12}
\begin{equation}\label{eq2}
P_{ap} = \frac{R-R_{ph}-R_d}{R_{ph}},
\end{equation}
where $R$ is the total count rate with laser illumination.

\section{Results and discussion}

\begin{figure}[tbp]
\centering
\includegraphics[width=\linewidth]{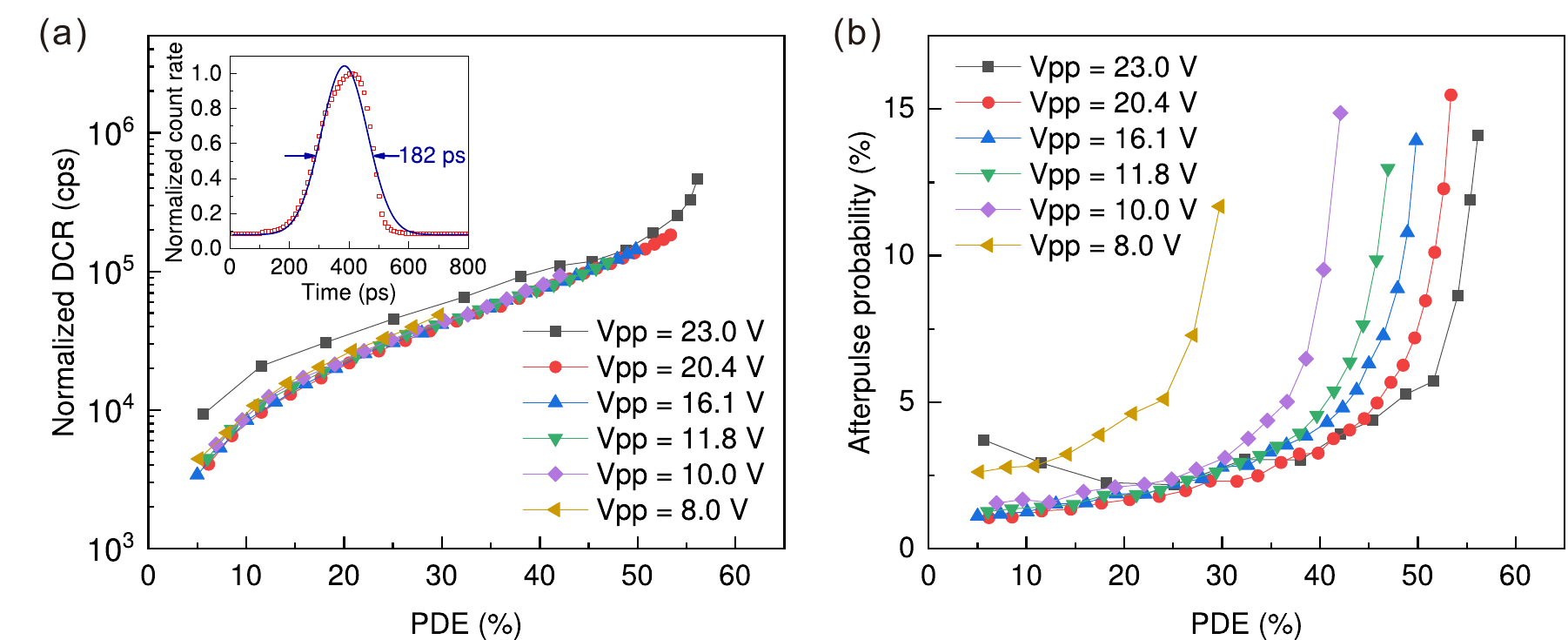}
\caption{Normalized DCR (a) and afterpulse probability (b) versus PDE of SPAD \#1 with gate amplitudes varied from 8.0 V to 23.0 V at 273 K. The inset plots the effective gating width measurement with 20.4 V $V_{pp}$ and 50\% PDE.}
\label{fig3}
\end{figure}

Two InGaAs/InP SPDs based on sample SPADs, i.e., SPAD \#1 and SPAD \#2, are characterized.
In the SWG scheme, gate amplitude and operation temperature are the most important factors for performance optimization.
First, SPAD \#1 is characterized at a fixed operation temperature to investigate the effect of gate amplitude.
Fig.~\ref{fig3}a plots the normalized DCR as a function of PDE with gate amplitudes varied from 8.0 V to 23.0 V at 273 K.
At each point, $t_w$ is measured by scanning the relative delay between laser pulses and sine wave gates.
For instance, the inset in Fig.~\ref{fig3}a shows a typically measured result and the fitting curve with 20.4 V $V_{pp}$ and 50\% PDE.
From Fig.~\ref{fig3}a, one can clearly observe that the lines of normalized DCR versus PDE with $V_{pp}$ from 8.0 V to 20.4 V overlap each other, which indicates that the normalized DCR-PDE trend is independent of the gate amplitude in such range.
The case with 23.0 V $V_{pp}$ is exceptional, and particularly in the region of $>$ 50\% PDE the normalized DCR is abnormally high.
This is probably due to the fact that gating width is too narrow to produce avalanches with enough signal-to-noise ratio in such case.

In addition, gate amplitude significantly affects the relationship between $P_{ap}$ and PDE, as illustrated in Fig.~\ref{fig3}b.
Given a certain PDE, $P_{ap}$ distinctly decreases as gate amplitude increases.
Such effect is due to that for the same excess bias $V_{ex}$, or PDE, higher gate amplitude results in smaller effective gating width and thus less charge carrier quantity of avalanche~\cite{Zhang15}.
As a consequence of the afterpulsing effect suppression, the maximum achievable PDE is also promoted with an increase of gate amplitude.

\begin{figure}[bp]
\centering
\includegraphics[width=\linewidth]{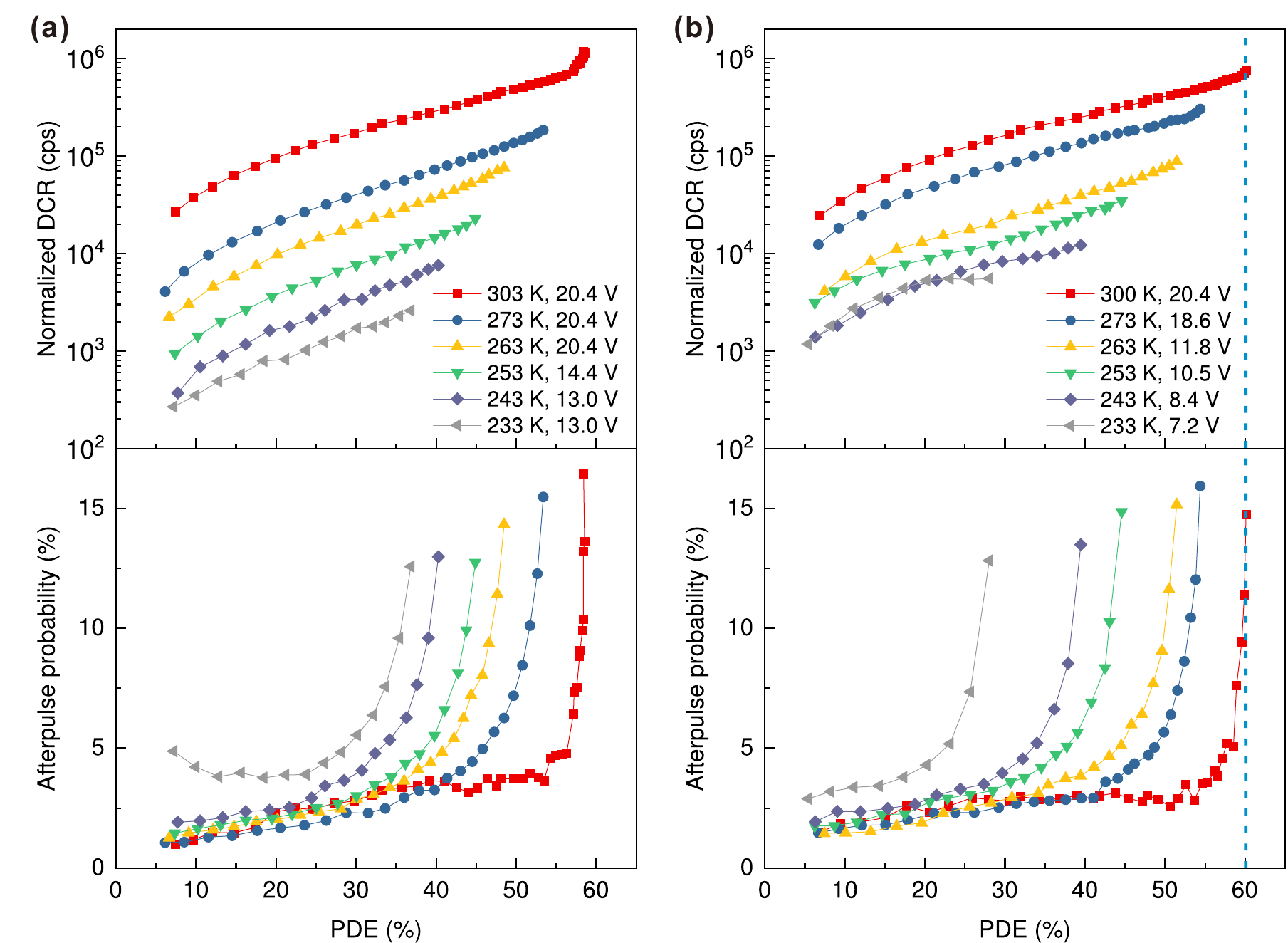}
\caption{Normalized DCR and afterpulse probability versus PDE of SPAD \#1 (a) and SPAD \#2 (b) with optimized gate amplitudes at different temperatures.}
\label{fig4}
\end{figure}

Then, we investigate the relationship between operation temperature and performance optimization.
Fig.~\ref{fig4}a and Fig.~\ref{fig4}b plot the parameters of normalized DCR and $P_{ap}$ as a function of PDE at different temperatures for SPAD \#1 and SPAD \#2, respectively.
At each temperature, the gate amplitude is optimized.
From Fig.~\ref{fig4}, one can observe that with an increase of operation temperature $P_{ap}$ apparently decreases given a fixed PDE, due to short lifetimes of trapped carriers at high temperatures.
More interestingly, the maximum achievable PDE also increases with temperature.
On one hand, the absorption coefficient of InGaAs and thus the absorption efficiency of incident photons increases with an increase of temperature~\cite{ZSS86}.
On the other hand, as the temperature increases the breakdown voltage and the gap between the breakdown voltage and the reach-through voltage of SPAD enlarge as well~\cite{LZC96}, which means that it is possible to apply enough high bias voltage to achieve the highest PDE only in the regime of high operation temperature.
At 303 K the maximum achievable PDE of SPAD \#1 reaches 58.4\% with 542 kcps DCR (1172 kcps normalized DCR) and 16.4\% $P_{ap}$, while at 300 K the maximum achievable PDE of SPAD \#2 reaches 60.1\% with 340 kcps DCR (744 kcps normalized DCR) and 14.8\% $P_{ap}$.
For practical use, the SPD exhibits significantly better performance than commercial products, e.g., $\sim$ 40\% PDE with 3 kcps DCR (14.5 kcps normalized DCR) and 5.5\% $P_{ap}$ at 253 K using SPAD \#2.

\section{Conclusion}

In conclusion, we have reported a high-frequency gating InGaAs/InP SPD with a PDE up to 60\% at 1550 nm.
By performing structure design optimization with a dielectric-metal reflection layer, the absorption efficiency of SPAD is significantly enhanced.
A monolithic readout circuit of avalanche extraction for 1.25 GHz SWG scheme is also designed to suppress the afterpulsing effect due to parasitic capacitance.
We have then optimized the conditions of gate amplitude and operating temperature to achieve a PDE as high as 60\%, with 340 kcps DCR and 14.8\% $P_{ap}$.
For practical use, the SPD exhibits $\sim$ 40\% PDE with 3 kcps DCR and 5.5\% $P_{ap}$, which can considerably improve the performance of diverse applications requiring InGaAs/InP SPDs for near-infrared single-photon detection.

\section*{acknowledgements}
This work has been supported by the National Key R\&D Program of China under Grant No.~2017YFA0304004, the National Natural Science Foundation of China under Grant No.~11674307, the Chinese Academy of Sciences, the Anhui Initiative in Quantum Information Technologies, and the Key Technology Innovation Project of Major Industry of Chongqing.

\bibliography{pde}

\begin{thebibliography}{39}%
\makeatletter
\providecommand \@ifxundefined [1]{%
 \@ifx{#1\undefined}
}%
\providecommand \@ifnum [1]{%
 \ifnum #1\expandafter \@firstoftwo
 \else \expandafter \@secondoftwo
 \fi
}%
\providecommand \@ifx [1]{%
 \ifx #1\expandafter \@firstoftwo
 \else \expandafter \@secondoftwo
 \fi
}%
\providecommand \natexlab [1]{#1}%
\providecommand \enquote  [1]{``#1''}%
\providecommand \bibnamefont  [1]{#1}%
\providecommand \bibfnamefont [1]{#1}%
\providecommand \citenamefont [1]{#1}%
\providecommand \href@noop [0]{\@secondoftwo}%
\providecommand \href [0]{\begingroup \@sanitize@url \@href}%
\providecommand \@href[1]{\@@startlink{#1}\@@href}%
\providecommand \@@href[1]{\endgroup#1\@@endlink}%
\providecommand \@sanitize@url [0]{\catcode `\\12\catcode `\$12\catcode
  `\&12\catcode `\#12\catcode `\^12\catcode `\_12\catcode `\%12\relax}%
\providecommand \@@startlink[1]{}%
\providecommand \@@endlink[0]{}%
\providecommand \url  [0]{\begingroup\@sanitize@url \@url }%
\providecommand \@url [1]{\endgroup\@href {#1}{\urlprefix }}%
\providecommand \urlprefix  [0]{URL }%
\providecommand \Eprint [0]{\href }%
\providecommand \doibase [0]{http://dx.doi.org/}%
\providecommand \selectlanguage [0]{\@gobble}%
\providecommand \bibinfo  [0]{\@secondoftwo}%
\providecommand \bibfield  [0]{\@secondoftwo}%
\providecommand \translation [1]{[#1]}%
\providecommand \BibitemOpen [0]{}%
\providecommand \bibitemStop [0]{}%
\providecommand \bibitemNoStop [0]{.\EOS\space}%
\providecommand \EOS [0]{\spacefactor3000\relax}%
\providecommand \BibitemShut  [1]{\csname bibitem#1\endcsname}%
\let\auto@bib@innerbib\@empty
\bibitem [{\citenamefont {Hadfield}(2009)}]{SPD09}%
  \BibitemOpen
  \bibfield  {author} {\bibinfo {author} {\bibfnamefont {R.~H.}\ \bibnamefont
  {Hadfield}},\ }\href {\doibase 10.1038/nphoton.2009.230} {\bibfield
  {journal} {\bibinfo  {journal} {Nat. Photonics}\ }\textbf {\bibinfo {volume}
  {3}},\ \bibinfo {pages} {696} (\bibinfo {year} {2009})}\BibitemShut {NoStop}%
\bibitem [{\citenamefont {Eisaman}\ \emph {et~al.}(2011)\citenamefont
  {Eisaman}, \citenamefont {Fan}, \citenamefont {Migdall},\ and\ \citenamefont
  {Polyakov}}]{SPD11}%
  \BibitemOpen
  \bibfield  {author} {\bibinfo {author} {\bibfnamefont {M.~D.}\ \bibnamefont
  {Eisaman}}, \bibinfo {author} {\bibfnamefont {J.}~\bibnamefont {Fan}},
  \bibinfo {author} {\bibfnamefont {A.}~\bibnamefont {Migdall}}, \ and\
  \bibinfo {author} {\bibfnamefont {S.~V.}\ \bibnamefont {Polyakov}},\ }\href
  {\doibase 10.1063/1.3610677} {\bibfield  {journal} {\bibinfo  {journal} {Rev.
  Sci. Instrum.}\ }\textbf {\bibinfo {volume} {82}},\ \bibinfo {pages} {071101}
  (\bibinfo {year} {2011})}\BibitemShut {NoStop}%
\bibitem [{\citenamefont {Gisin}\ \emph {et~al.}(2002)\citenamefont {Gisin},
  \citenamefont {Ribordy}, \citenamefont {Tittel},\ and\ \citenamefont
  {Zbinden}}]{QKD02}%
  \BibitemOpen
  \bibfield  {author} {\bibinfo {author} {\bibfnamefont {N.}~\bibnamefont
  {Gisin}}, \bibinfo {author} {\bibfnamefont {G.}~\bibnamefont {Ribordy}},
  \bibinfo {author} {\bibfnamefont {W.}~\bibnamefont {Tittel}}, \ and\ \bibinfo
  {author} {\bibfnamefont {H.}~\bibnamefont {Zbinden}},\ }\href {\doibase
  10.1103/RevModPhys.74.145} {\bibfield  {journal} {\bibinfo  {journal} {Rev.
  Mod. Phys.}\ }\textbf {\bibinfo {volume} {74}},\ \bibinfo {pages} {145}
  (\bibinfo {year} {2002})}\BibitemShut {NoStop}%
\bibitem [{\citenamefont {Yu}\ \emph {et~al.}(2017)\citenamefont {Yu},
  \citenamefont {Shangguan}, \citenamefont {Xia}, \citenamefont {Zhang},
  \citenamefont {Dou},\ and\ \citenamefont {Pan}}]{Yu17}%
  \BibitemOpen
  \bibfield  {author} {\bibinfo {author} {\bibfnamefont {C.}~\bibnamefont
  {Yu}}, \bibinfo {author} {\bibfnamefont {M.}~\bibnamefont {Shangguan}},
  \bibinfo {author} {\bibfnamefont {H.}~\bibnamefont {Xia}}, \bibinfo {author}
  {\bibfnamefont {J.}~\bibnamefont {Zhang}}, \bibinfo {author} {\bibfnamefont
  {X.}~\bibnamefont {Dou}}, \ and\ \bibinfo {author} {\bibfnamefont {J.-W.}\
  \bibnamefont {Pan}},\ }\href {\doibase 10.1364/OE.25.014611} {\bibfield
  {journal} {\bibinfo  {journal} {Opt. Express}\ }\textbf {\bibinfo {volume}
  {25}},\ \bibinfo {pages} {14611} (\bibinfo {year} {2017})}\BibitemShut
  {NoStop}%
\bibitem [{\citenamefont {Eraerds}\ \emph {et~al.}(2010)\citenamefont
  {Eraerds}, \citenamefont {Legr\'{e}}, \citenamefont {Zhang}, \citenamefont
  {Zbinden},\ and\ \citenamefont {Gisin}}]{OTDR10}%
  \BibitemOpen
  \bibfield  {author} {\bibinfo {author} {\bibfnamefont {P.}~\bibnamefont
  {Eraerds}}, \bibinfo {author} {\bibfnamefont {M.}~\bibnamefont {Legr\'{e}}},
  \bibinfo {author} {\bibfnamefont {J.}~\bibnamefont {Zhang}}, \bibinfo
  {author} {\bibfnamefont {H.}~\bibnamefont {Zbinden}}, \ and\ \bibinfo
  {author} {\bibfnamefont {N.}~\bibnamefont {Gisin}},\ }\href
  {http://jlt.osa.org/abstract.cfm?URI=jlt-28-6-952} {\bibfield  {journal}
  {\bibinfo  {journal} {J. Lightwave Technol.}\ }\textbf {\bibinfo {volume}
  {28}},\ \bibinfo {pages} {952} (\bibinfo {year} {2010})}\BibitemShut
  {NoStop}%
\bibitem [{\citenamefont {Marsili}\ \emph {et~al.}(2013)\citenamefont
  {Marsili}, \citenamefont {Verma}, \citenamefont {Stern}, \citenamefont
  {Harrington}, \citenamefont {Lita}, \citenamefont {Gerrits}, \citenamefont
  {Vayshenker}, \citenamefont {Baek}, \citenamefont {Shaw}, \citenamefont
  {Mirin},\ and\ \citenamefont {Nam}}]{SNSPD13}%
  \BibitemOpen
  \bibfield  {author} {\bibinfo {author} {\bibfnamefont {F.}~\bibnamefont
  {Marsili}}, \bibinfo {author} {\bibfnamefont {V.~B.}\ \bibnamefont {Verma}},
  \bibinfo {author} {\bibfnamefont {J.~A.}\ \bibnamefont {Stern}}, \bibinfo
  {author} {\bibfnamefont {S.}~\bibnamefont {Harrington}}, \bibinfo {author}
  {\bibfnamefont {A.~E.}\ \bibnamefont {Lita}}, \bibinfo {author}
  {\bibfnamefont {T.}~\bibnamefont {Gerrits}}, \bibinfo {author} {\bibfnamefont
  {I.}~\bibnamefont {Vayshenker}}, \bibinfo {author} {\bibfnamefont
  {B.}~\bibnamefont {Baek}}, \bibinfo {author} {\bibfnamefont {M.~D.}\
  \bibnamefont {Shaw}}, \bibinfo {author} {\bibfnamefont {R.~P.}\ \bibnamefont
  {Mirin}}, \ and\ \bibinfo {author} {\bibfnamefont {S.~W.}\ \bibnamefont
  {Nam}},\ }\href {https://doi.org/10.1038/nphoton.2013.13} {\bibfield
  {journal} {\bibinfo  {journal} {Nat. Photonics}\ }\textbf {\bibinfo {volume}
  {7}},\ \bibinfo {pages} {210} (\bibinfo {year} {2013})}\BibitemShut {NoStop}%
\bibitem [{\citenamefont {Zhang}\ \emph {et~al.}(2017)\citenamefont {Zhang},
  \citenamefont {You}, \citenamefont {Li}, \citenamefont {Huang}, \citenamefont
  {Lv}, \citenamefont {Zhang}, \citenamefont {Liu}, \citenamefont {Wu},
  \citenamefont {Wang},\ and\ \citenamefont {Xie}}]{SNSPD17}%
  \BibitemOpen
  \bibfield  {author} {\bibinfo {author} {\bibfnamefont {W.}~\bibnamefont
  {Zhang}}, \bibinfo {author} {\bibfnamefont {L.}~\bibnamefont {You}}, \bibinfo
  {author} {\bibfnamefont {H.}~\bibnamefont {Li}}, \bibinfo {author}
  {\bibfnamefont {J.}~\bibnamefont {Huang}}, \bibinfo {author} {\bibfnamefont
  {C.}~\bibnamefont {Lv}}, \bibinfo {author} {\bibfnamefont {L.}~\bibnamefont
  {Zhang}}, \bibinfo {author} {\bibfnamefont {X.}~\bibnamefont {Liu}}, \bibinfo
  {author} {\bibfnamefont {J.}~\bibnamefont {Wu}}, \bibinfo {author}
  {\bibfnamefont {Z.}~\bibnamefont {Wang}}, \ and\ \bibinfo {author}
  {\bibfnamefont {X.}~\bibnamefont {Xie}},\ }\href {\doibase
  10.1007/s11433-017-9113-4} {\bibfield  {journal} {\bibinfo  {journal}
  {Science China Physics, Mechanics {\&} Astronomy}\ }\textbf {\bibinfo
  {volume} {60}},\ \bibinfo {pages} {120314} (\bibinfo {year}
  {2017})}\BibitemShut {NoStop}%
\bibitem [{\citenamefont {Langrock}\ \emph {et~al.}(2005)\citenamefont
  {Langrock}, \citenamefont {Diamanti}, \citenamefont {Roussev}, \citenamefont
  {Yamamoto}, \citenamefont {Fejer},\ and\ \citenamefont {Takesue}}]{UCSPD05}%
  \BibitemOpen
  \bibfield  {author} {\bibinfo {author} {\bibfnamefont {C.}~\bibnamefont
  {Langrock}}, \bibinfo {author} {\bibfnamefont {E.}~\bibnamefont {Diamanti}},
  \bibinfo {author} {\bibfnamefont {R.~V.}\ \bibnamefont {Roussev}}, \bibinfo
  {author} {\bibfnamefont {Y.}~\bibnamefont {Yamamoto}}, \bibinfo {author}
  {\bibfnamefont {M.~M.}\ \bibnamefont {Fejer}}, \ and\ \bibinfo {author}
  {\bibfnamefont {H.}~\bibnamefont {Takesue}},\ }\href {\doibase
  10.1364/OL.30.001725} {\bibfield  {journal} {\bibinfo  {journal} {Opt.
  Lett.}\ }\textbf {\bibinfo {volume} {30}},\ \bibinfo {pages} {1725} (\bibinfo
  {year} {2005})}\BibitemShut {NoStop}%
\bibitem [{\citenamefont {Shentu}\ \emph {et~al.}(2013)\citenamefont {Shentu},
  \citenamefont {Pelc}, \citenamefont {Wang}, \citenamefont {Sun},
  \citenamefont {Zheng}, \citenamefont {Fejer}, \citenamefont {Zhang},\ and\
  \citenamefont {Pan}}]{Shentu13}%
  \BibitemOpen
  \bibfield  {author} {\bibinfo {author} {\bibfnamefont {G.-L.}\ \bibnamefont
  {Shentu}}, \bibinfo {author} {\bibfnamefont {J.~S.}\ \bibnamefont {Pelc}},
  \bibinfo {author} {\bibfnamefont {X.-D.}\ \bibnamefont {Wang}}, \bibinfo
  {author} {\bibfnamefont {Q.-C.}\ \bibnamefont {Sun}}, \bibinfo {author}
  {\bibfnamefont {M.-Y.}\ \bibnamefont {Zheng}}, \bibinfo {author}
  {\bibfnamefont {M.~M.}\ \bibnamefont {Fejer}}, \bibinfo {author}
  {\bibfnamefont {Q.}~\bibnamefont {Zhang}}, \ and\ \bibinfo {author}
  {\bibfnamefont {J.-W.}\ \bibnamefont {Pan}},\ }\href {\doibase
  10.1364/OE.21.013986} {\bibfield  {journal} {\bibinfo  {journal} {Opt.
  Express}\ }\textbf {\bibinfo {volume} {21}},\ \bibinfo {pages} {13986}
  (\bibinfo {year} {2013})}\BibitemShut {NoStop}%
\bibitem [{\citenamefont {Tosi}\ \emph {et~al.}(2009)\citenamefont {Tosi},
  \citenamefont {Mora}, \citenamefont {Zappa},\ and\ \citenamefont
  {Cova}}]{TMZ09}%
  \BibitemOpen
  \bibfield  {author} {\bibinfo {author} {\bibfnamefont {A.}~\bibnamefont
  {Tosi}}, \bibinfo {author} {\bibfnamefont {A.~D.}\ \bibnamefont {Mora}},
  \bibinfo {author} {\bibfnamefont {F.}~\bibnamefont {Zappa}}, \ and\ \bibinfo
  {author} {\bibfnamefont {S.}~\bibnamefont {Cova}},\ }\href {\doibase
  10.1080/09500340802263075} {\bibfield  {journal} {\bibinfo  {journal} {J.
  Mod. Opt.}\ }\textbf {\bibinfo {volume} {56}},\ \bibinfo {pages} {299}
  (\bibinfo {year} {2009})}\BibitemShut {NoStop}%
\bibitem [{\citenamefont {Itzler}\ \emph {et~al.}(2011)\citenamefont {Itzler},
  \citenamefont {Jiang}, \citenamefont {Entwistle}, \citenamefont {Slomkowski},
  \citenamefont {Tosi}, \citenamefont {Acerbi}, \citenamefont {Zappa},\ and\
  \citenamefont {Cova}}]{IJE11}%
  \BibitemOpen
  \bibfield  {author} {\bibinfo {author} {\bibfnamefont {M.~A.}\ \bibnamefont
  {Itzler}}, \bibinfo {author} {\bibfnamefont {X.}~\bibnamefont {Jiang}},
  \bibinfo {author} {\bibfnamefont {M.}~\bibnamefont {Entwistle}}, \bibinfo
  {author} {\bibfnamefont {K.}~\bibnamefont {Slomkowski}}, \bibinfo {author}
  {\bibfnamefont {A.}~\bibnamefont {Tosi}}, \bibinfo {author} {\bibfnamefont
  {F.}~\bibnamefont {Acerbi}}, \bibinfo {author} {\bibfnamefont
  {F.}~\bibnamefont {Zappa}}, \ and\ \bibinfo {author} {\bibfnamefont
  {S.}~\bibnamefont {Cova}},\ }\href {\doibase 10.1080/09500340.2010.547262}
  {\bibfield  {journal} {\bibinfo  {journal} {J. Mod. Opt.}\ }\textbf {\bibinfo
  {volume} {58}},\ \bibinfo {pages} {174} (\bibinfo {year} {2011})}\BibitemShut
  {NoStop}%
\bibitem [{\citenamefont {Zhang}\ \emph {et~al.}(2015)\citenamefont {Zhang},
  \citenamefont {Itzler}, \citenamefont {Zbinden},\ and\ \citenamefont
  {Pan}}]{Zhang15}%
  \BibitemOpen
  \bibfield  {author} {\bibinfo {author} {\bibfnamefont {J.}~\bibnamefont
  {Zhang}}, \bibinfo {author} {\bibfnamefont {M.~A.}\ \bibnamefont {Itzler}},
  \bibinfo {author} {\bibfnamefont {H.}~\bibnamefont {Zbinden}}, \ and\
  \bibinfo {author} {\bibfnamefont {J.-W.}\ \bibnamefont {Pan}},\ }\href
  {\doibase 10.1038/lsa.2015.59} {\bibfield  {journal} {\bibinfo  {journal}
  {Light: Science \& Applications}\ }\textbf {\bibinfo {volume} {4}},\ \bibinfo
  {pages} {e286} (\bibinfo {year} {2015})}\BibitemShut {NoStop}%
\bibitem [{IDQ()}]{IDQ}%
  \BibitemOpen
  \href@noop {} {}\bibinfo {howpublished}
  {\url{https://www.idquantique.com/quantum-sensing/products/id221/}}\BibitemShut
  {NoStop}%
\bibitem [{CTe()}]{CTek}%
  \BibitemOpen
  \href@noop {} {}\bibinfo {howpublished}
  {\url{http://www.quantum-info.com/English/product/pfour/danguangzitanceqi/2017/0831/303.html}}\BibitemShut
  {NoStop}%
\bibitem [{\citenamefont {Jiang}\ \emph {et~al.}(2018)\citenamefont {Jiang},
  \citenamefont {Gao}, \citenamefont {Fang}, \citenamefont {Liu}, \citenamefont
  {Zhou}, \citenamefont {Jiang}, \citenamefont {Chen}, \citenamefont {Jin},
  \citenamefont {Zhang},\ and\ \citenamefont {Pan}}]{Jiang18}%
  \BibitemOpen
  \bibfield  {author} {\bibinfo {author} {\bibfnamefont {W.-H.}\ \bibnamefont
  {Jiang}}, \bibinfo {author} {\bibfnamefont {X.-J.}\ \bibnamefont {Gao}},
  \bibinfo {author} {\bibfnamefont {Y.-Q.}\ \bibnamefont {Fang}}, \bibinfo
  {author} {\bibfnamefont {J.-H.}\ \bibnamefont {Liu}}, \bibinfo {author}
  {\bibfnamefont {Y.}~\bibnamefont {Zhou}}, \bibinfo {author} {\bibfnamefont
  {L.-Q.}\ \bibnamefont {Jiang}}, \bibinfo {author} {\bibfnamefont
  {W.}~\bibnamefont {Chen}}, \bibinfo {author} {\bibfnamefont {G.}~\bibnamefont
  {Jin}}, \bibinfo {author} {\bibfnamefont {J.}~\bibnamefont {Zhang}}, \ and\
  \bibinfo {author} {\bibfnamefont {J.-W.}\ \bibnamefont {Pan}},\ }\href
  {\doibase 10.1063/1.5055376} {\bibfield  {journal} {\bibinfo  {journal} {Rev.
  Sci. Instrum.}\ }\textbf {\bibinfo {volume} {89}},\ \bibinfo {pages} {123104}
  (\bibinfo {year} {2018})}\BibitemShut {NoStop}%
\bibitem [{\citenamefont {Rarity}\ \emph {et~al.}(2000)\citenamefont {Rarity},
  \citenamefont {Wall}, \citenamefont {Ridley}, \citenamefont {Owens},\ and\
  \citenamefont {Tapster}}]{Rarity00}%
  \BibitemOpen
  \bibfield  {author} {\bibinfo {author} {\bibfnamefont {J.~G.}\ \bibnamefont
  {Rarity}}, \bibinfo {author} {\bibfnamefont {T.~E.}\ \bibnamefont {Wall}},
  \bibinfo {author} {\bibfnamefont {K.~D.}\ \bibnamefont {Ridley}}, \bibinfo
  {author} {\bibfnamefont {P.~C.~M.}\ \bibnamefont {Owens}}, \ and\ \bibinfo
  {author} {\bibfnamefont {P.~R.}\ \bibnamefont {Tapster}},\ }\href {\doibase
  10.1364/AO.39.006746} {\bibfield  {journal} {\bibinfo  {journal} {Appl.
  Opt.}\ }\textbf {\bibinfo {volume} {39}},\ \bibinfo {pages} {6746} (\bibinfo
  {year} {2000})}\BibitemShut {NoStop}%
\bibitem [{\citenamefont {Thew}\ \emph {et~al.}(2007)\citenamefont {Thew},
  \citenamefont {Stucki}, \citenamefont {Gautier}, \citenamefont {Zbinden},\
  and\ \citenamefont {Rochas}}]{TSG07}%
  \BibitemOpen
  \bibfield  {author} {\bibinfo {author} {\bibfnamefont {R.~T.}\ \bibnamefont
  {Thew}}, \bibinfo {author} {\bibfnamefont {D.}~\bibnamefont {Stucki}},
  \bibinfo {author} {\bibfnamefont {J.-D.}\ \bibnamefont {Gautier}}, \bibinfo
  {author} {\bibfnamefont {H.}~\bibnamefont {Zbinden}}, \ and\ \bibinfo
  {author} {\bibfnamefont {A.}~\bibnamefont {Rochas}},\ }\href {\doibase
  10.1063/1.2815916} {\bibfield  {journal} {\bibinfo  {journal} {Appl. Phys.
  Lett.}\ }\textbf {\bibinfo {volume} {91}},\ \bibinfo {pages} {201114}
  (\bibinfo {year} {2007})}\BibitemShut {NoStop}%
\bibitem [{\citenamefont {{Liu}}\ \emph {et~al.}(2008)\citenamefont {{Liu}},
  \citenamefont {{Hu}}, \citenamefont {{Campbell}}, \citenamefont {{Pan}},\
  and\ \citenamefont {{Tashima}}}]{Liu08}%
  \BibitemOpen
  \bibfield  {author} {\bibinfo {author} {\bibfnamefont {M.}~\bibnamefont
  {{Liu}}}, \bibinfo {author} {\bibfnamefont {C.}~\bibnamefont {{Hu}}},
  \bibinfo {author} {\bibfnamefont {J.~C.}\ \bibnamefont {{Campbell}}},
  \bibinfo {author} {\bibfnamefont {Z.}~\bibnamefont {{Pan}}}, \ and\ \bibinfo
  {author} {\bibfnamefont {M.~M.}\ \bibnamefont {{Tashima}}},\ }\href {\doibase
  10.1109/JQE.2007.916688} {\bibfield  {journal} {\bibinfo  {journal} {IEEE J.
  Quantum Electron.}\ }\textbf {\bibinfo {volume} {44}},\ \bibinfo {pages}
  {430} (\bibinfo {year} {2008})}\BibitemShut {NoStop}%
\bibitem [{\citenamefont {{Zhang}}\ \emph {et~al.}(2009)\citenamefont
  {{Zhang}}, \citenamefont {{Thew}}, \citenamefont {{Gautier}}, \citenamefont
  {{Gisin}},\ and\ \citenamefont {{Zbinden}}}]{Zhang09}%
  \BibitemOpen
  \bibfield  {author} {\bibinfo {author} {\bibfnamefont {J.}~\bibnamefont
  {{Zhang}}}, \bibinfo {author} {\bibfnamefont {R.}~\bibnamefont {{Thew}}},
  \bibinfo {author} {\bibfnamefont {J.}~\bibnamefont {{Gautier}}}, \bibinfo
  {author} {\bibfnamefont {N.}~\bibnamefont {{Gisin}}}, \ and\ \bibinfo
  {author} {\bibfnamefont {H.}~\bibnamefont {{Zbinden}}},\ }\href {\doibase
  10.1109/JQE.2009.2013210} {\bibfield  {journal} {\bibinfo  {journal} {IEEE J.
  Quantum Electron.}\ }\textbf {\bibinfo {volume} {45}},\ \bibinfo {pages}
  {792} (\bibinfo {year} {2009})}\BibitemShut {NoStop}%
\bibitem [{\citenamefont {Warburton}, \citenamefont {Itzler},\ and\
  \citenamefont {Buller}(2009)}]{WIB09}%
  \BibitemOpen
  \bibfield  {author} {\bibinfo {author} {\bibfnamefont {R.~E.}\ \bibnamefont
  {Warburton}}, \bibinfo {author} {\bibfnamefont {M.}~\bibnamefont {Itzler}}, \
  and\ \bibinfo {author} {\bibfnamefont {G.~S.}\ \bibnamefont {Buller}},\
  }\href {\doibase 10.1063/1.3079668} {\bibfield  {journal} {\bibinfo
  {journal} {Appl. Phys. Lett.}\ }\textbf {\bibinfo {volume} {94}},\ \bibinfo
  {pages} {071116} (\bibinfo {year} {2009})}\BibitemShut {NoStop}%
\bibitem [{\citenamefont {Itzler}\ \emph {et~al.}(2009)\citenamefont {Itzler},
  \citenamefont {Jiang}, \citenamefont {Nyman},\ and\ \citenamefont
  {Slomkowski}}]{Itzler09}%
  \BibitemOpen
  \bibfield  {author} {\bibinfo {author} {\bibfnamefont {M.~A.}\ \bibnamefont
  {Itzler}}, \bibinfo {author} {\bibfnamefont {X.}~\bibnamefont {Jiang}},
  \bibinfo {author} {\bibfnamefont {B.}~\bibnamefont {Nyman}}, \ and\ \bibinfo
  {author} {\bibfnamefont {K.}~\bibnamefont {Slomkowski}},\ }\bibfield
  {booktitle} {\emph {\bibinfo {booktitle} {Quantum Sensing and Nanophotonic
  Devices VI}},\ }\href {\doibase 10.1117/12.814669} {\bibfield  {journal}
  {\bibinfo  {journal} {Proc. SPIE}\ }\textbf {\bibinfo {volume} {7222}},\
  \bibinfo {pages} {462 } (\bibinfo {year} {2009})}\BibitemShut {NoStop}%
\bibitem [{\citenamefont {Yan}\ \emph {et~al.}(2012)\citenamefont {Yan},
  \citenamefont {Hamel}, \citenamefont {Heinrichs}, \citenamefont {Jiang},
  \citenamefont {Itzler},\ and\ \citenamefont {Jennewein}}]{Yan12}%
  \BibitemOpen
  \bibfield  {author} {\bibinfo {author} {\bibfnamefont {Z.}~\bibnamefont
  {Yan}}, \bibinfo {author} {\bibfnamefont {D.~R.}\ \bibnamefont {Hamel}},
  \bibinfo {author} {\bibfnamefont {A.~K.}\ \bibnamefont {Heinrichs}}, \bibinfo
  {author} {\bibfnamefont {X.}~\bibnamefont {Jiang}}, \bibinfo {author}
  {\bibfnamefont {M.~A.}\ \bibnamefont {Itzler}}, \ and\ \bibinfo {author}
  {\bibfnamefont {T.}~\bibnamefont {Jennewein}},\ }\href {\doibase
  10.1063/1.4732813} {\bibfield  {journal} {\bibinfo  {journal} {Rev. Sci.
  Instrum.}\ }\textbf {\bibinfo {volume} {83}},\ \bibinfo {pages} {073105}
  (\bibinfo {year} {2012})}\BibitemShut {NoStop}%
\bibitem [{\citenamefont {Korzh}\ \emph {et~al.}(2014)\citenamefont {Korzh},
  \citenamefont {Walenta}, \citenamefont {Lunghi}, \citenamefont {Gisin},\ and\
  \citenamefont {Zbinden}}]{KWL14}%
  \BibitemOpen
  \bibfield  {author} {\bibinfo {author} {\bibfnamefont {B.}~\bibnamefont
  {Korzh}}, \bibinfo {author} {\bibfnamefont {N.}~\bibnamefont {Walenta}},
  \bibinfo {author} {\bibfnamefont {T.}~\bibnamefont {Lunghi}}, \bibinfo
  {author} {\bibfnamefont {N.}~\bibnamefont {Gisin}}, \ and\ \bibinfo {author}
  {\bibfnamefont {H.}~\bibnamefont {Zbinden}},\ }\href {\doibase
  10.1063/1.4866582} {\bibfield  {journal} {\bibinfo  {journal} {Appl. Phys.
  Lett.}\ }\textbf {\bibinfo {volume} {104}},\ \bibinfo {pages} {081108}
  (\bibinfo {year} {2014})}\BibitemShut {NoStop}%
\bibitem [{\citenamefont {Yu}\ \emph {et~al.}(2018)\citenamefont {Yu},
  \citenamefont {Qiu}, \citenamefont {Xia}, \citenamefont {Dou}, \citenamefont
  {Zhang},\ and\ \citenamefont {Pan}}]{Yu18}%
  \BibitemOpen
  \bibfield  {author} {\bibinfo {author} {\bibfnamefont {C.}~\bibnamefont
  {Yu}}, \bibinfo {author} {\bibfnamefont {J.}~\bibnamefont {Qiu}}, \bibinfo
  {author} {\bibfnamefont {H.}~\bibnamefont {Xia}}, \bibinfo {author}
  {\bibfnamefont {X.}~\bibnamefont {Dou}}, \bibinfo {author} {\bibfnamefont
  {J.}~\bibnamefont {Zhang}}, \ and\ \bibinfo {author} {\bibfnamefont {J.-W.}\
  \bibnamefont {Pan}},\ }\href {\doibase 10.1063/1.5047472} {\bibfield
  {journal} {\bibinfo  {journal} {Rev. Sci. Instrum.}\ }\textbf {\bibinfo
  {volume} {89}},\ \bibinfo {pages} {103106} (\bibinfo {year}
  {2018})}\BibitemShut {NoStop}%
\bibitem [{\citenamefont {Yuan}\ \emph {et~al.}(2007)\citenamefont {Yuan},
  \citenamefont {Kardynal}, \citenamefont {Sharpe},\ and\ \citenamefont
  {Shields}}]{SD07}%
  \BibitemOpen
  \bibfield  {author} {\bibinfo {author} {\bibfnamefont {Z.~L.}\ \bibnamefont
  {Yuan}}, \bibinfo {author} {\bibfnamefont {B.~E.}\ \bibnamefont {Kardynal}},
  \bibinfo {author} {\bibfnamefont {A.~W.}\ \bibnamefont {Sharpe}}, \ and\
  \bibinfo {author} {\bibfnamefont {A.~J.}\ \bibnamefont {Shields}},\ }\href
  {\doibase 10.1063/1.2760135} {\bibfield  {journal} {\bibinfo  {journal}
  {Appl. Phys. Lett.}\ }\textbf {\bibinfo {volume} {91}},\ \bibinfo {pages}
  {041114} (\bibinfo {year} {2007})}\BibitemShut {NoStop}%
\bibitem [{\citenamefont {Xu}\ \emph {et~al.}(2009)\citenamefont {Xu},
  \citenamefont {Wu}, \citenamefont {Gu}, \citenamefont {Jian}, \citenamefont
  {Wu},\ and\ \citenamefont {Zeng}}]{SD09}%
  \BibitemOpen
  \bibfield  {author} {\bibinfo {author} {\bibfnamefont {L.}~\bibnamefont
  {Xu}}, \bibinfo {author} {\bibfnamefont {E.}~\bibnamefont {Wu}}, \bibinfo
  {author} {\bibfnamefont {X.}~\bibnamefont {Gu}}, \bibinfo {author}
  {\bibfnamefont {Y.}~\bibnamefont {Jian}}, \bibinfo {author} {\bibfnamefont
  {G.}~\bibnamefont {Wu}}, \ and\ \bibinfo {author} {\bibfnamefont
  {H.}~\bibnamefont {Zeng}},\ }\href {\doibase 10.1063/1.3120224} {\bibfield
  {journal} {\bibinfo  {journal} {Appl. Phys. Lett.}\ }\textbf {\bibinfo
  {volume} {94}},\ \bibinfo {pages} {161106} (\bibinfo {year}
  {2009})}\BibitemShut {NoStop}%
\bibitem [{\citenamefont {Yuan}\ \emph {et~al.}(2010)\citenamefont {Yuan},
  \citenamefont {Sharpe}, \citenamefont {Dynes}, \citenamefont {Dixon},\ and\
  \citenamefont {Shields}}]{SD10}%
  \BibitemOpen
  \bibfield  {author} {\bibinfo {author} {\bibfnamefont {Z.~L.}\ \bibnamefont
  {Yuan}}, \bibinfo {author} {\bibfnamefont {A.~W.}\ \bibnamefont {Sharpe}},
  \bibinfo {author} {\bibfnamefont {J.~F.}\ \bibnamefont {Dynes}}, \bibinfo
  {author} {\bibfnamefont {A.~R.}\ \bibnamefont {Dixon}}, \ and\ \bibinfo
  {author} {\bibfnamefont {A.~J.}\ \bibnamefont {Shields}},\ }\href {\doibase
  10.1063/1.3309698} {\bibfield  {journal} {\bibinfo  {journal} {Appl. Phys.
  Lett.}\ }\textbf {\bibinfo {volume} {96}},\ \bibinfo {pages} {071101}
  (\bibinfo {year} {2010})}\BibitemShut {NoStop}%
\bibitem [{\citenamefont {Comandar}\ \emph {et~al.}(2015)\citenamefont
  {Comandar}, \citenamefont {Fr\"ohlich}, \citenamefont {Dynes}, \citenamefont
  {Sharpe}, \citenamefont {Lucamarini}, \citenamefont {Yuan}, \citenamefont
  {Penty},\ and\ \citenamefont {Shields}}]{SD15}%
  \BibitemOpen
  \bibfield  {author} {\bibinfo {author} {\bibfnamefont {L.~C.}\ \bibnamefont
  {Comandar}}, \bibinfo {author} {\bibfnamefont {B.}~\bibnamefont
  {Fr\"ohlich}}, \bibinfo {author} {\bibfnamefont {J.~F.}\ \bibnamefont
  {Dynes}}, \bibinfo {author} {\bibfnamefont {A.~W.}\ \bibnamefont {Sharpe}},
  \bibinfo {author} {\bibfnamefont {M.}~\bibnamefont {Lucamarini}}, \bibinfo
  {author} {\bibfnamefont {Z.~L.}\ \bibnamefont {Yuan}}, \bibinfo {author}
  {\bibfnamefont {R.~V.}\ \bibnamefont {Penty}}, \ and\ \bibinfo {author}
  {\bibfnamefont {A.~J.}\ \bibnamefont {Shields}},\ }\href {\doibase
  10.1063/1.4913527} {\bibfield  {journal} {\bibinfo  {journal} {J. Appl.
  Phys.}\ }\textbf {\bibinfo {volume} {117}},\ \bibinfo {pages} {083109}
  (\bibinfo {year} {2015})}\BibitemShut {NoStop}%
\bibitem [{\citenamefont {Namekata}, \citenamefont {Sasamori},\ and\
  \citenamefont {Inoue}(2006)}]{SWG06}%
  \BibitemOpen
  \bibfield  {author} {\bibinfo {author} {\bibfnamefont {N.}~\bibnamefont
  {Namekata}}, \bibinfo {author} {\bibfnamefont {S.}~\bibnamefont {Sasamori}},
  \ and\ \bibinfo {author} {\bibfnamefont {S.}~\bibnamefont {Inoue}},\ }\href
  {\doibase 10.1364/OE.14.010043} {\bibfield  {journal} {\bibinfo  {journal}
  {Opt. Express}\ }\textbf {\bibinfo {volume} {14}},\ \bibinfo {pages} {10043}
  (\bibinfo {year} {2006})}\BibitemShut {NoStop}%
\bibitem [{\citenamefont {Zhang}\ \emph {et~al.}(2009)\citenamefont {Zhang},
  \citenamefont {Thew}, \citenamefont {Barreiro},\ and\ \citenamefont
  {Zbinden}}]{Zhang09-2}%
  \BibitemOpen
  \bibfield  {author} {\bibinfo {author} {\bibfnamefont {J.}~\bibnamefont
  {Zhang}}, \bibinfo {author} {\bibfnamefont {R.}~\bibnamefont {Thew}},
  \bibinfo {author} {\bibfnamefont {C.}~\bibnamefont {Barreiro}}, \ and\
  \bibinfo {author} {\bibfnamefont {H.}~\bibnamefont {Zbinden}},\ }\href
  {\doibase 10.1063/1.3223576} {\bibfield  {journal} {\bibinfo  {journal}
  {Appl. Phys. Lett.}\ }\textbf {\bibinfo {volume} {95}},\ \bibinfo {pages}
  {091103} (\bibinfo {year} {2009})}\BibitemShut {NoStop}%
\bibitem [{\citenamefont {Liang}\ \emph {et~al.}(2012)\citenamefont {Liang},
  \citenamefont {Liu}, \citenamefont {Wang}, \citenamefont {Du}, \citenamefont
  {Ma}, \citenamefont {Jin}, \citenamefont {Chen}, \citenamefont {Zhang},\ and\
  \citenamefont {Pan}}]{Liang12}%
  \BibitemOpen
  \bibfield  {author} {\bibinfo {author} {\bibfnamefont {X.-L.}\ \bibnamefont
  {Liang}}, \bibinfo {author} {\bibfnamefont {J.-H.}\ \bibnamefont {Liu}},
  \bibinfo {author} {\bibfnamefont {Q.}~\bibnamefont {Wang}}, \bibinfo {author}
  {\bibfnamefont {D.-B.}\ \bibnamefont {Du}}, \bibinfo {author} {\bibfnamefont
  {J.}~\bibnamefont {Ma}}, \bibinfo {author} {\bibfnamefont {G.}~\bibnamefont
  {Jin}}, \bibinfo {author} {\bibfnamefont {Z.-B.}\ \bibnamefont {Chen}},
  \bibinfo {author} {\bibfnamefont {J.}~\bibnamefont {Zhang}}, \ and\ \bibinfo
  {author} {\bibfnamefont {J.-W.}\ \bibnamefont {Pan}},\ }\href {\doibase
  10.1063/1.4746291} {\bibfield  {journal} {\bibinfo  {journal} {Rev. Sci.
  Instrum.}\ }\textbf {\bibinfo {volume} {83}},\ \bibinfo {pages} {083111}
  (\bibinfo {year} {2012})}\BibitemShut {NoStop}%
\bibitem [{\citenamefont {Walenta}\ \emph {et~al.}(2012)\citenamefont
  {Walenta}, \citenamefont {Lunghi}, \citenamefont {Guinnard}, \citenamefont
  {Houlmann}, \citenamefont {Zbinden},\ and\ \citenamefont {Gisin}}]{SWG12}%
  \BibitemOpen
  \bibfield  {author} {\bibinfo {author} {\bibfnamefont {N.}~\bibnamefont
  {Walenta}}, \bibinfo {author} {\bibfnamefont {T.}~\bibnamefont {Lunghi}},
  \bibinfo {author} {\bibfnamefont {O.}~\bibnamefont {Guinnard}}, \bibinfo
  {author} {\bibfnamefont {R.}~\bibnamefont {Houlmann}}, \bibinfo {author}
  {\bibfnamefont {H.}~\bibnamefont {Zbinden}}, \ and\ \bibinfo {author}
  {\bibfnamefont {N.}~\bibnamefont {Gisin}},\ }\href {\doibase
  10.1063/1.4749802} {\bibfield  {journal} {\bibinfo  {journal} {J. Appl.
  Phys.}\ }\textbf {\bibinfo {volume} {112}},\ \bibinfo {pages} {063106}
  (\bibinfo {year} {2012})}\BibitemShut {NoStop}%
\bibitem [{\citenamefont {Restelli}, \citenamefont {Bienfang},\ and\
  \citenamefont {Migdall}(2013)}]{SWG13}%
  \BibitemOpen
  \bibfield  {author} {\bibinfo {author} {\bibfnamefont {A.}~\bibnamefont
  {Restelli}}, \bibinfo {author} {\bibfnamefont {J.~C.}\ \bibnamefont
  {Bienfang}}, \ and\ \bibinfo {author} {\bibfnamefont {A.~L.}\ \bibnamefont
  {Migdall}},\ }\href {\doibase 10.1063/1.4801939} {\bibfield  {journal}
  {\bibinfo  {journal} {Appl. Phys. Lett.}\ }\textbf {\bibinfo {volume}
  {102}},\ \bibinfo {pages} {141104} (\bibinfo {year} {2013})}\BibitemShut
  {NoStop}%
\bibitem [{\citenamefont {Jiang}\ \emph {et~al.}(2017)\citenamefont {Jiang},
  \citenamefont {Liu}, \citenamefont {Liu}, \citenamefont {Jin}, \citenamefont
  {Zhang},\ and\ \citenamefont {Pan}}]{Jiang17}%
  \BibitemOpen
  \bibfield  {author} {\bibinfo {author} {\bibfnamefont {W.-H.}\ \bibnamefont
  {Jiang}}, \bibinfo {author} {\bibfnamefont {J.-H.}\ \bibnamefont {Liu}},
  \bibinfo {author} {\bibfnamefont {Y.}~\bibnamefont {Liu}}, \bibinfo {author}
  {\bibfnamefont {G.}~\bibnamefont {Jin}}, \bibinfo {author} {\bibfnamefont
  {J.}~\bibnamefont {Zhang}}, \ and\ \bibinfo {author} {\bibfnamefont {J.-W.}\
  \bibnamefont {Pan}},\ }\href {\doibase 10.1364/OL.42.005090} {\bibfield
  {journal} {\bibinfo  {journal} {Opt. Lett.}\ }\textbf {\bibinfo {volume}
  {42}},\ \bibinfo {pages} {5090} (\bibinfo {year} {2017})}\BibitemShut
  {NoStop}%
\bibitem [{\citenamefont {Ma}\ \emph {et~al.}(2016)\citenamefont {Ma},
  \citenamefont {Bai}, \citenamefont {Wang}, \citenamefont {Tong},
  \citenamefont {Jin}, \citenamefont {Zhang},\ and\ \citenamefont
  {Pan}}]{Ma16}%
  \BibitemOpen
  \bibfield  {author} {\bibinfo {author} {\bibfnamefont {J.}~\bibnamefont
  {Ma}}, \bibinfo {author} {\bibfnamefont {B.}~\bibnamefont {Bai}}, \bibinfo
  {author} {\bibfnamefont {L.-J.}\ \bibnamefont {Wang}}, \bibinfo {author}
  {\bibfnamefont {C.-Z.}\ \bibnamefont {Tong}}, \bibinfo {author}
  {\bibfnamefont {G.}~\bibnamefont {Jin}}, \bibinfo {author} {\bibfnamefont
  {J.}~\bibnamefont {Zhang}}, \ and\ \bibinfo {author} {\bibfnamefont {J.-W.}\
  \bibnamefont {Pan}},\ }\href {\doibase 10.1364/AO.55.007497} {\bibfield
  {journal} {\bibinfo  {journal} {Appl. Opt.}\ }\textbf {\bibinfo {volume}
  {55}},\ \bibinfo {pages} {7497} (\bibinfo {year} {2016})}\BibitemShut
  {NoStop}%
\bibitem [{\citenamefont {Itzler}\ \emph {et~al.}(2007)\citenamefont {Itzler},
  \citenamefont {r.~Ben-Michael}, \citenamefont {Hsu}, \citenamefont
  {Slomkowski}, \citenamefont {Tosi}, \citenamefont {Cova}, \citenamefont
  {Zappa},\ and\ \citenamefont {Ispasoiu}}]{IBH07}%
  \BibitemOpen
  \bibfield  {author} {\bibinfo {author} {\bibfnamefont {M.~A.}\ \bibnamefont
  {Itzler}}, \bibinfo {author} {\bibnamefont {r.~Ben-Michael}}, \bibinfo
  {author} {\bibfnamefont {C.~F.}\ \bibnamefont {Hsu}}, \bibinfo {author}
  {\bibfnamefont {K.}~\bibnamefont {Slomkowski}}, \bibinfo {author}
  {\bibfnamefont {A.}~\bibnamefont {Tosi}}, \bibinfo {author} {\bibfnamefont
  {S.}~\bibnamefont {Cova}}, \bibinfo {author} {\bibfnamefont {F.}~\bibnamefont
  {Zappa}}, \ and\ \bibinfo {author} {\bibfnamefont {R.}~\bibnamefont
  {Ispasoiu}},\ }\href {\doibase 10.1080/09500340600792291} {\bibfield
  {journal} {\bibinfo  {journal} {J. Mod. Opt.}\ }\textbf {\bibinfo {volume}
  {54}},\ \bibinfo {pages} {283} (\bibinfo {year} {2007})}\BibitemShut
  {NoStop}%
\bibitem [{\citenamefont {Lee}\ \emph {et~al.}(2009)\citenamefont {Lee},
  \citenamefont {Ban}, \citenamefont {Makino}, \citenamefont {Hayashi},
  \citenamefont {Nagatsuma}, \citenamefont {Mita}, \citenamefont {Tanaka},
  \citenamefont {Matsuoka}, \citenamefont {Sugawara}, \citenamefont {Tsuji},
  \citenamefont {Aoki}, \citenamefont {Takamatsu}, \citenamefont {Sasada},
  \citenamefont {Yamamoto}, \citenamefont {Okayasu}, \citenamefont {Toyonaka},
  \citenamefont {Inoue},\ and\ \citenamefont {Shishikura}}]{LBM09}%
  \BibitemOpen
  \bibfield  {author} {\bibinfo {author} {\bibfnamefont {Y.}~\bibnamefont
  {Lee}}, \bibinfo {author} {\bibfnamefont {T.}~\bibnamefont {Ban}}, \bibinfo
  {author} {\bibfnamefont {S.}~\bibnamefont {Makino}}, \bibinfo {author}
  {\bibfnamefont {H.}~\bibnamefont {Hayashi}}, \bibinfo {author} {\bibfnamefont
  {K.}~\bibnamefont {Nagatsuma}}, \bibinfo {author} {\bibfnamefont
  {R.}~\bibnamefont {Mita}}, \bibinfo {author} {\bibfnamefont {S.}~\bibnamefont
  {Tanaka}}, \bibinfo {author} {\bibfnamefont {Y.}~\bibnamefont {Matsuoka}},
  \bibinfo {author} {\bibfnamefont {T.}~\bibnamefont {Sugawara}}, \bibinfo
  {author} {\bibfnamefont {S.}~\bibnamefont {Tsuji}}, \bibinfo {author}
  {\bibfnamefont {M.}~\bibnamefont {Aoki}}, \bibinfo {author} {\bibfnamefont
  {H.}~\bibnamefont {Takamatsu}}, \bibinfo {author} {\bibfnamefont
  {M.}~\bibnamefont {Sasada}}, \bibinfo {author} {\bibfnamefont
  {H.}~\bibnamefont {Yamamoto}}, \bibinfo {author} {\bibfnamefont
  {M.}~\bibnamefont {Okayasu}}, \bibinfo {author} {\bibfnamefont
  {T.}~\bibnamefont {Toyonaka}}, \bibinfo {author} {\bibfnamefont
  {H.}~\bibnamefont {Inoue}}, \ and\ \bibinfo {author} {\bibfnamefont
  {M.}~\bibnamefont {Shishikura}},\ }in\ \href
  {http://www.osapublishing.org/abstract.cfm?URI=OFC-2009-JThA28} {\emph
  {\bibinfo {booktitle} {Optical Fiber Communication Conference and National
  Fiber Optic Engineers Conference}}}\ (\bibinfo  {publisher} {Optical Society
  of America},\ \bibinfo {year} {2009})\ p.\ \bibinfo {pages}
  {JThA28}\BibitemShut {NoStop}%
\bibitem [{\citenamefont {Zielinski}\ \emph {et~al.}(1986)\citenamefont
  {Zielinski}, \citenamefont {Schweizer}, \citenamefont {Streubel},
  \citenamefont {Eisele},\ and\ \citenamefont {Weimann}}]{ZSS86}%
  \BibitemOpen
  \bibfield  {author} {\bibinfo {author} {\bibfnamefont {E.}~\bibnamefont
  {Zielinski}}, \bibinfo {author} {\bibfnamefont {H.}~\bibnamefont
  {Schweizer}}, \bibinfo {author} {\bibfnamefont {K.}~\bibnamefont {Streubel}},
  \bibinfo {author} {\bibfnamefont {H.}~\bibnamefont {Eisele}}, \ and\ \bibinfo
  {author} {\bibfnamefont {G.}~\bibnamefont {Weimann}},\ }\href {\doibase
  10.1063/1.336358} {\bibfield  {journal} {\bibinfo  {journal} {J. Appl.
  Phys.}\ }\textbf {\bibinfo {volume} {59}},\ \bibinfo {pages} {2196} (\bibinfo
  {year} {1986})}\BibitemShut {NoStop}%
\bibitem [{\citenamefont {Lacaita}\ \emph {et~al.}(1996)\citenamefont
  {Lacaita}, \citenamefont {Zappa}, \citenamefont {Cova},\ and\ \citenamefont
  {Lovati}}]{LZC96}%
  \BibitemOpen
  \bibfield  {author} {\bibinfo {author} {\bibfnamefont {A.}~\bibnamefont
  {Lacaita}}, \bibinfo {author} {\bibfnamefont {F.}~\bibnamefont {Zappa}},
  \bibinfo {author} {\bibfnamefont {S.}~\bibnamefont {Cova}}, \ and\ \bibinfo
  {author} {\bibfnamefont {P.}~\bibnamefont {Lovati}},\ }\href {\doibase
  10.1364/AO.35.002986} {\bibfield  {journal} {\bibinfo  {journal} {Appl.
  Opt.}\ }\textbf {\bibinfo {volume} {35}},\ \bibinfo {pages} {2986} (\bibinfo
  {year} {1996})}\BibitemShut {NoStop}%
\end{thebibliography}%

\end{document}